\documentclass[a4paper, amsfonts, amssymb, amsmath, reprint, showkeys, footinbib, twoside,superscriptaddress,floatfix,longbibliography]{revtex4-1}
\usepackage{tabularx}
\usepackage{booktabs}
\usepackage{placeins}
\usepackage{multirow} 
\usepackage{graphicx}%
\usepackage{dcolumn}%
\usepackage{bm}%
\usepackage{float}
\usepackage{xcolor}
\usepackage{mhchem}
\usepackage{gensymb}
\usepackage[slantedGreek]{newtxmath}
\usepackage{comment}
\usepackage[pdftex, bookmarks, pdffitwindow=false, pdfstartview=FitH, pdfdisplaydoctitle, colorlinks, plainpages=false, pdftitle={},pdfauthor={}, pdfpagelabels, hypertexnames, citecolor={blue!50!black},linkcolor={blue!50!black}, urlcolor={blue!50!black}, pdflang={en}, hyperfootnotes=false, breaklinks]{hyperref}
\usepackage{siunitx}
\usepackage{subcaption}
\usepackage{graphicx}
\usepackage[normalem]{ulem}

\newcommand{\avg}[1]{\ensuremath{\left<#1\right>}}

\newcommand{\figLabelCapt}[1]{\textbf{\MakeLowercase{{#1}}}}
\newcommand{\refSub}[2]{\hyperref[#2]{\ref{#2}\figLabelCapt{#1}}}

\newcommand{\figrefsub}[2]{Fig.~\refSub{#2}{#1}}

\usepackage[justification=raggedright,singlelinecheck=false]{caption}

\makeatletter
\def\@bibdataout@aps{
 \immediate\write\@bibdataout{
 @CONTROL{
   apsrev41Control, author="48",editor="1",pages="0",title="0",year="1"
 }}
 \if@filesw
  \immediate\write\@auxout{\string\citation{apsrev41Control}}
 \fi
}
\makeatother

\begin{document}

\title{Chemical potentials from structure factors: I. Neutral multi-component mixtures}

\author{Roya Savoj}
\thanks{These authors contributed equally.}
\affiliation{Department of Chemistry, UC Berkeley, California 94720, United States}

\author{Xiaoyu Wang}
\thanks{These authors contributed equally.}
\affiliation{Department of Chemistry, UC Berkeley, California 94720, United States}

\author{Musahid Ahmed}
\affiliation{Chemical Sciences Division, Lawrence Berkeley National Laboratory, Berkeley, California, 94720, United States}

\author{Bingqing Cheng}
\email{bingqingcheng@berkeley.edu}
\affiliation{Department of Chemistry, UC Berkeley, California 94720, United States}
\affiliation{Chemical Sciences Division, Lawrence Berkeley National Laboratory, Berkeley, California, 94720, United States}
\affiliation{Bakar Institute of Digital Materials for the Planet, UC Berkeley, California 94720, United States}

\date{\today}

\begin{abstract}
The chemical potentials of multi-component mixtures underlie many physical and chemical phenomena, but remain challenging to compute. 
The S0 method enables the computation of chemical potentials from equilibrium molecular dynamics simulations, by leveraging the thermodynamic relationship between particle number fluctuations and derivatives of chemical potentials, followed by numerical integration along different compositions. 
Here we generalize the S0 method from two-component mixtures to neutral multi-component mixtures.
We first extend the statistical mechanical formalism to high-dimensional compositional space, and then introduce a Gaussian process integration scheme combined with active learning to efficiently integrate chemical potentials and sample diverse compositions.
We use this method to compute the mixing free energies of a molten metal alloy, and the solubilities of two paracetamol polymorphs in water-ethanol solvents.
The extended S0 method provides a practical and scalable route for computing chemical potentials in neutral bulk multi-component mixtures from atomistic simulations.
\end{abstract}

\maketitle

\date{\today}%

\section{Introduction}

The chemical potential of each component in a mixture is central to a wide range of physical and chemical phenomena, including solvation, osmosis, phase equilibria, and phase transitions. 
However, obtaining these chemical potentials from atomistic simulations is challenging using standard free energy methods~\cite{sanz2007solubility,paluch2010method,lisal2005molecular,moucka2011molecular,perego2016chemical,joung2008determination,Li2017,Li2018,Vinutha2021}.
Monte Carlo particle insertion or removal methods~\cite{allen2012computer,smit1989calculation} often suffer from poor convergence, particularly in condensed phases or for large molecules~\cite{perego2016chemical}. 
Thermodynamic integration and overlapping distribution approaches~\cite{sanz2007solubility,paluch2010method} mitigate some of these convergence issues, but require multiple simulations across intermediate thermodynamic states or coupling parameters~\cite{Li2017,Li2018}. 
Kirkwood–Buff integrals evaluated in real space~\cite{Kirkwood1951,dawass2019kirkwood} suffer from significant finite-size effects, necessitating careful extrapolation schemes~\cite{CortesHuerto2016,dawass2019kirkwood}.

The S0 method~\cite{Cheng2022Computing} was developed to compute the composition dependence of chemical potentials from equilibrium molecular dynamics (MD) simulations using modest system sizes.
This method is based on the thermodynamic relationship between composition fluctuations and derivatives of the chemical potentials with respect to concentration~\cite{Kirkwood1951}. In practice, it only requires static structure factors obtained from equilibrium isothermal–isobaric (NPT) MD simulations at different solute concentrations, and the use of numerical integration over concentration to obtain the chemical potentials. 
Originally, this method was formulated to predict chemical potential derivatives for two-component solutions. The S0 method has been utilized in several thermodynamic applications, such as predicting the solubility of molecular crystals~\cite{reinhardt2023streamlined, Herboth2025Estimating}, determining phase equilibria in mixtures~\cite{Cheng2023Diamond,Wang2026hydrogen}, computing adsorption isotherms for confined fluids~\cite{Schmid2023Computing} and the formation of azeotropes in binary mixtures~\cite{Wang2024Integrating}.

Here, we generalize the S0 method to multi-component bulk mixtures. 
We first extend the statistical mechanical framework to treat an arbitrary high-dimensional compositional space. 
To compute chemical potentials efficiently and robustly, we employ an integration scheme based on Gaussian process regression, combined with active learning strategies for sampling additional compositions to simulate. 
In this paper, we focus on systems where each component is neutral so long-range electrostatic correlations are unimportant in the S0 calculation of chemical potentials.
We will follow up with a Part II paper that specifically treats ionic mixtures.
In what follows, we first present the methodology, followed by two representative applications to molten metal alloys and paracetamol in water-ethanol solvents.

\twocolumngrid
\section{Theory}
\subsection{Particle-number fluctuations and structure factors}
We consider a bulk multi-component system in the grand canonical ensemble $(\{\mu\} = \{\mu_1, \mu_2, \ldots, \mu_C\}, V, T)$ with $C$ types of particles.  
Its grand potential $\Omega$ can be expressed as
\begin{equation}
\Omega \left( \{ \mu \}, V, T \right)  = -k_\mathrm{B} T \ln
\sum_{\{ n \}}
\exp \left( \sum_{i = 1}^C \dfrac{n_i \mu_i}{k_\mathrm{B} T} \right) 
Q\left( \{ n \}, V, T \right) ,
\label{eq:grandPotential}
\end{equation}
where $\{n\} = \{n_1, n_2, \ldots, n_C\}$ denotes the number of particles for each type,
and $Q$ is the canonical partition function for the system at the fixed $\{ n\}$, temperature $T$, and volume $V$. 

Take two of the components that we label $\alpha$ and $\beta$,
the second-order derivative of $\Omega$ with respect to their chemical potentials is \cite{Cheng2022Computing,molecularBen-naim2006,Kirkwood1951}: 
\begin{equation}
   -\dfrac{\partial^2 \Omega}{\partial \mu_{\alpha} \partial \mu_{\beta}} 
   = \dfrac{\partial \langle n_{\alpha} \rangle_{\{ \mu \},V,T}}{\partial \mu_{\beta}} 
   = \dfrac{1}{k_{\rm B} T}\langle \Delta n_{\alpha} \Delta n_{\beta} \rangle_{\{ \mu \},V,T}, 
   \label{eq:particlefluc2ndder}
\end{equation}
where $\langle \cdots \rangle$ indicates an ensemble average, and 
$\Delta n_{\alpha} = n_{\alpha} - \langle n_{\alpha} \rangle_{\{ \mu \},V,T}$ is the particle-number fluctuation.
Eqn.~\eqref{eq:particlefluc2ndder} means that particle-number fluctuations directly yield the response of the mean particle number to changes in chemical potentials in the grand canonical ensemble. 
Such response can be written in a matrix component form as
\begin{equation}
        B_{\alpha \beta} = 
        \dfrac{k_{\rm B}T}{\avg{N}}
        \left(\dfrac{\partial \avg{ n_{\alpha} }}
        {\partial \mu_{\beta}}\right)_{V, T},
    \label{eq:MatrixB}
\end{equation}
where $N = \sum_{i=1}^C n_i$.

As derived in Ref.~\cite{Cheng2022Computing}, particle-number fluctuations are related to the partial static structure factors approaching the zero-wavevector limit:
\begin{equation}
    S_{\alpha \beta}^0 = 
    \lim_{ k \rightarrow 0} S_{\alpha \beta}(\mathbf{k}) 
    = \dfrac{\langle
    \Delta n_{\alpha} \Delta n_{\beta} \rangle_{\{ \mu \},V,T} }
    {\langle N \rangle \sqrt{x_\alpha x_\beta}}
    = \dfrac{1}{\sqrt{x_\alpha x_\beta}}B_{\alpha \beta}
    \label{eq:k0lim},
\end{equation}
where $x_\alpha = \avg{n_\alpha} / \avg{N} $ denotes the particle number fraction, and $k=|\mathbf{k}|$.
Indeed, the elements of $\mathbf{B}$ can also be written in terms of the real-space Kirkwood--Buff integrals in the thermodynamic limit $G_{\alpha\beta}^{\infty}$~\cite{Connell1971Thermodynamic, Fingerhut2020Thermodynamic}: 
\begin{equation}
B_{\alpha\beta}
= x_\alpha\delta_{\alpha\beta}
+ \rho x_\alpha x_\beta G_{\alpha\beta}^{\infty},
\label{eq:B_KBI}
\end{equation}
where $\rho$ is the total number density and $\delta_{\alpha\beta}$ is the Kronecker delta.

Moreover, the structure factors can be computed from molecular dynamics simulations in the isothermal-isobaric ($\{ n \},P,T$) ensemble as
\begin{equation}
        S_{\alpha \beta}(\mathbf{k}) = \dfrac{1}{\sqrt{n_\alpha n_\beta}}
        \avg{\rho_\alpha(\mathbf{k})\rho_\beta(-\mathbf{k})}_{\{ n \},P,T},
        \label{eq:sfsim}
\end{equation}
where
\begin{equation}
\rho_\alpha(\mathbf{k}) 
= \sum_{j=1}^{n_\alpha} \exp(i \mathbf{k} \cdot \mathbf{r}_{j_\alpha}),
\label{eq:rhokt}
\end{equation}
and $\mathbf{r}_{j_\alpha}$ denotes the position of particle type $\alpha$ with index $j$.

To extrapolate to $S_{\alpha \beta}^0$,
Ref.~\cite{Cheng2022Computing} used the Ornstein–Zernike (OZ) form for neutral mixtures,
\begin{equation}
    S_{\alpha \beta}^{\mathrm{OZ}}(\mathbf k) = 
    \dfrac{S_{\alpha \beta}^0}{1+a k^2},
    \label{eq:oz}
\end{equation}
where $a$ is treated as a fitting parameter.
Here we introduce a generalized, matrix OZ form to fit the whole $\mathbf{S}$ matrix:
\begin{equation}
    \mathbf{S}^{\mathrm{OZ}}(\mathbf{k})
    =
    \left[
    \mathbf{S}^{-1}(0)
    +
    k^2\mathbf{L}
    \right]^{-1},
    \label{eq:matrix_oz}
\end{equation}
performed by optimizing symmetric matrices $\mathbf{S}(0)$ and $\mathbf{L}$ simultaneously to all diagonal and off-diagonal elements of $\mathbf{S}(\mathbf{k})$.

\subsection{Chemical potential derivatives}
Ultimately, we want to compute the chemical potential of each component in a multi-component system as a function of composition at constant temperature and pressure. 
This requires evaluating chemical potential derivatives in the ($\{n\},P,T$) ensemble,
\begin{equation}
  U_{\alpha \beta} =  N \left(\dfrac{\partial \mu_{\alpha}}{\partial n_{\beta}}\right)_{n_i \neq n_\beta, P,T}.
  \label{eq:U}
\end{equation}
In Appendix Sec.~\ref{sec:ensembleSwitching} we derive the relation between the $\mathbf{B}$ and the $\mathbf{U}$ matrices, following the established thermodynamic transformations with matrix formulations based on Kirkwood–Buff theory ~\cite{Kirkwood1951, molecularBen-naim2006, Nichols2009Improved, robustBusselez2025, Connell1971Thermodynamic}:

\begin{equation}
\mathbf{U} = k_{\rm B}T\left[\mathbf{B}^{-1} -
\dfrac{\mathbf{B}^{-1}\mathbf{x}\mathbf{x}^T\mathbf{B}^{-1}}{\mathbf{x}^T\mathbf{B}^{-1}\mathbf{x}}\right],
\label{eq:muder}
\end{equation}
where $\mathbf{x} = [x_1, x_2, \ldots, x_C]^T$. 
The first term on the right-hand side of Eqn.~\eqref{eq:muder} is how the chemical potential changes with particle number at fixed volume, and the second term accounts for the difference between fixed pressure and fixed volume conditions.
The partial derivative matrix $\mathbf{U}$ satisfies the Gibbs–Duhem condition, $\mathbf{x}^T\mathbf{U} = \mathbf{0}^T$.

\subsection{Total derivative of chemical potential}

To arrive at the absolute $\mu$ by integration using the partial derivatives in the $\mathbf{U}$ matrix (Eqn.~\eqref{eq:muder}),
it is essential to note that the particle number fractions, $\mathbf{x}$, are constrained.
In a general case, there are $f=C-1$ independent variables as $\sum_{i=1}^C x_i=1$; additional constraints, such as charge neutrality can further reduce the number of degrees of freedom, which we will discuss in the follow-up manuscript.
To produce a set of $f$ independent variables, $\tilde{\mathbf{x}}$,
we introduce a projection operation:
\begin{equation}
\mathbf{Q} \mathbf{x} =
\begin{bmatrix}
\mathbf{P}_{f \times C} \\ \hline
\mathbf{1}_{1\times C} 
\end{bmatrix}
\mathbf{x}
=
\begin{bmatrix}
\tilde{\mathbf{x}} \\ \hline
1 
\end{bmatrix},
\label{eq:selectionMat}
\end{equation}
and we note in passing that additional constraints can be easily incorporated into this projection scheme.
There is substantial freedom in the choice of the projection matrix $\mathbf{P}$, as long as it is of rank $f$.
The simplest choice is to set all the diagonal elements  
$P_{ii}$ to 1 and all the other elements to 0.
Consequently, the full set of particle number fractions, $\mathbf{x}$ can be reconstructed from $\tilde{\mathbf{x}}$, using
\begin{equation}
    \mathbf{Q}^{-1}
    \begin{bmatrix}
    \tilde{\mathbf{x}} \\ \hline
        1 
    \end{bmatrix}
= \mathbf{x}.
\end{equation}

We then take matrix $\mathbf{M}$ to be the first $f$ columns of $\mathbf{Q}^{-1}$, 
$\mathbf{M} = (\mathbf{Q}^{-1})_{[:,1:f]}$, so
\begin{equation}
    \dfrac{\partial x_i}{\partial \widetilde{x}_j} = \mathbf{M}_{ij}.
    \label{eq:x2xtilde}
\end{equation}
Next, we define matrix $\mathbf{\Gamma}$ with elements
\begin{multline}
    \Gamma_{\alpha\beta} = \left(\dfrac{\partial \mu_\alpha}{\partial \tilde{x}_\beta}\right)_{\tilde{x} \neq \tilde{x}_\beta } 
    =
    \sum_{i=1}^{C}\left(\dfrac{\partial \mu_{\alpha}}{\partial n_{i}}\right) \left(\dfrac{\partial n_i}{\partial \tilde{x}_{\beta}}\right)_{\tilde{x} \neq \tilde{x}_{\beta}}\\
    =
    \sum_{i=1}^{C} N \left(\dfrac{\partial \mu_\alpha}{\partial n_i}\right)\left(\dfrac{\partial x_i}{\partial \tilde{x}_\beta}\right)_{\tilde{x} \neq \tilde{x}_{\beta}} +
    \sum_{i=1}^{C}\frac{n_i}{N}\left(\dfrac{\partial \mu_\alpha}{\partial n_i}\right) \left(\dfrac{dN}{d\tilde{x}_\beta}\right)_{\tilde{x} \neq \tilde{x}_{\beta}},
    \label{eq:gammaAB}
\end{multline}
where the second term in the last line sums to zero~\cite{callen2006thermodynamics}.
Putting together Eqns.~\eqref{eq:gammaAB} and~\eqref{eq:x2xtilde},
\begin{equation}
    \mathbf{\Gamma} = \mathbf{U}\mathbf{M}.
    \label{eq:gamma}
\end{equation}
The Gibbs–Duhem condition is preserved, $\mathbf{x}^T\mathbf{\Gamma} = \mathbf{0}^T$.
Finally, the total derivative of chemical potential of component $\alpha$ is
\begin{equation}
    d\mu_{\alpha} =\sum_{i=1}^f \Gamma_{\alpha i} d\tilde{x}_i,
    \label{eq:diffequi}
\end{equation}
from which one can integrate to obtain $\mu_{\alpha}$.
In practice, for better numerical behavior, we often operate on the excess chemical potentials
\begin{equation}
    \mu^{\mathrm{ex}}_{\alpha} = \mu_{\alpha} - k_{\rm B} T\ln(x_\alpha),
    \label{eq:mu_ex}
\end{equation}
by integrating the partial derivatives
 \begin{equation}
      \Gamma^{\mathrm{ex}}_{\alpha \beta} 
      = \Gamma_{\alpha\beta} -k_{\rm B}T \frac{1}{x_{\alpha}} M_{\alpha \beta}.
      \label{eq:gammaexeq}
 \end{equation}

 \subsection{Gaussian Process Integration}
 \label{sec:GP}

From MD simulations at a set of independent compositions $\tilde{\mathbf{x}}$, 
the derivative matrices $\mathbf{\Gamma}$s can be obtained from structure factors via Eqns.~\eqref{eq:k0lim},~\eqref{eq:muder}, and~\eqref{eq:gamma}.
To determine the chemical potential $\mu_{\alpha}(\tilde{\mathbf{x}})$, one can use Eqn.~\eqref{eq:diffequi} and a reference point
to perform numerical integration along a specific one-dimensional path in the compositional space.
However, such estimated $\mu_{\alpha}(\tilde{\mathbf{x}})$ suffer from path-dependent statistical errors and numerical errors in the integrand.
Moreover, for determining $\mu_{\alpha}(\tilde{\mathbf{x}})$ over the whole compositional space of $\tilde{\mathbf{x}}$, it is unclear how to efficiently select the integration paths.

\begin{figure}[H]
    \centering
    \includegraphics[width=1.0\linewidth]{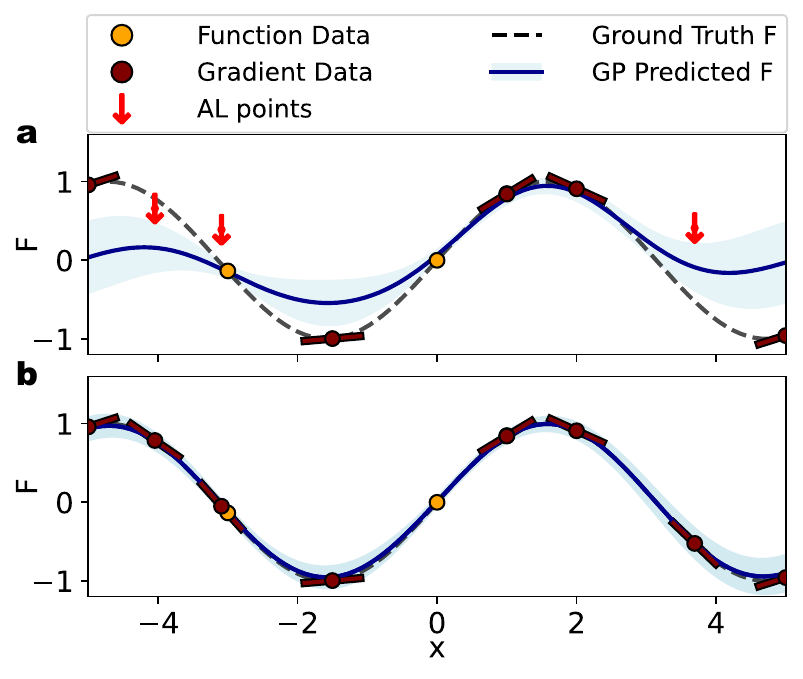}
    \caption{Gaussian process (GP) regression from function and gradient data. 
    \figLabelCapt{a} GP prediction of $\mathrm{sin}(x)$ from the initial function and gradient observations, with the uncertainty estimates shown as the shaded area.
    Red arrows indicate new gradients picked by active learning (AL). 
    \figLabelCapt{b} GP prediction after adding the AL points.}
    \label{fig:GPandALschematic}
\end{figure}

To overcome such difficulties, we employ Gaussian process (GP) regression~\cite{Miryashkin2023Bayesian, Deringer2021Gaussian, Mones2016Exploration}, which enables a global reconstruction of $\mu_\alpha(\tilde{\mathbf{x}})$.
While the most standard GP fits to observed function values~\cite{williams2006gaussian}, here we adopt the formulation that can incorporate function values and gradient information simultaneously.
As an illustration,
the upper panel of Fig.~\ref{fig:GPandALschematic} shows GP regression using both noisy function and gradient data from a ground truth function.

In the general setting, given observations at inputs $\mathbf{X}$, consisting of values $\mathbf{Y}$ at $\mathbf{X}_f$, and gradient values $\nabla\mathbf{Y}$ at $\mathbf{X}_g$,
the GP posterior mean and covariance at new input locations $\mathbf{X}_*$ are written as
\begin{equation}
    \tilde{\mathbf{Y}}_* =  
    K_{\mathbf{X}_* \mathbf{X}}
    \left(K_{\mathbf{X} \mathbf{X}} + \Sigma\right)^{-1}  
    \begin{bmatrix}
        \mathbf{Y} \\\nabla \mathbf{Y}
    \end{bmatrix},
    \label{eq:GP}
\end{equation}
and 
\begin{equation}
    \mathrm{cov}(\tilde{\mathbf{Y}}_*) = 
    K_{\mathbf{X}_* \mathbf{X}_*} 
    - K_{\mathbf{X}_* \mathbf{X}}
\left(K_{\mathbf{X} \mathbf{X}} + \Sigma\right)^{-1}  
    K_{\mathbf{X} \mathbf{X}_*}  .
    \label{eq:covarianceGP}
\end{equation}
The noise matrix has a block-diagonal form
\begin{equation}
    \Sigma  =
    \begin{bmatrix}
    \sigma_f^2\mathbf{I} 
    & \mathbf{0} \\
    \mathbf{0}
    & \sigma_g^2\mathbf{I} \\
    \end{bmatrix},
\end{equation}
where $\sigma_f$ and $\sigma_g$ denote the noise levels of the measured function data and gradients.
Elements of the kernel matrix K involving derivatives are obtained by differentiating the base kernel function $k$.
For example, the block of the kernel matrix for data vector $[Y, \nabla_{1} Y]$ at points $[X, X']$ is
\begin{equation}
        \begin{bmatrix}
k(X,X') & \dfrac{\partial}{\partial X_1'} k(X,X') \\
\dfrac{\partial}{\partial X_1} k(X,X') & \dfrac{\partial^2}{\partial X_1\partial X_1'} k(X,X')
    \end{bmatrix}.
\end{equation}

Many choices of the kernel functions $k$ are available. 
A commonly-used stationary kernel is the radial basis function (RBF), defined as:
\begin{equation}
   k_\mathrm{RBF}(X,X') = \exp\left(-\dfrac{||X-X'||^2}{2\theta^2}\right),
\label{eq:RBF_kernel}
\end{equation}
where $\theta$ is the fixed length scale parameter and can be anisotropic for each input dimension.
On the other hand, for data with spatially varying changes in correlation length,
non-stationary kernels with adaptive covariance that depends on the input location can be advantageous~\cite{williams2006gaussian}.
One simple approach is to warp the inputs before applying a stationary kernel~\cite{sampson1992nonparametric,snoek2014input}, e.g.
the $d$-th dimension of the input, $X_d$, can be transformed as
\begin{equation}
w_d(X_d)=\frac{\exp(\alpha_d X_d)-1}{\exp(\alpha_d)-1}, 
\label{eq:input_warp}
\end{equation}
where $\alpha_d$ controls the strength of the nonlinear warping along the dimension $d$, and $\alpha_d\to 0$ gives $w_d(X_d)=X_d$ that recovers the stationary kernel. 
Other popular non-stationary kernels include the Gibbs kernel~\cite{gibbs1997bayesian,paciorek2004nonstationary}, which makes the length scale an explicit function of the input $\mathbf{X}$,
and the non-stationary Mat\'ern kernel~\cite{paciorek2006spatial,remes2017non}, which combines an input-dependent length scale with an adjustable degree of differentiability.
The hyperparameters (e.g., $\sigma_f$, $\sigma_g$, $\theta$) of the kernels and of the GP can be set a priori based on the underlying problem,
or optimized by maximizing the log marginal likelihood~\cite{williams2006gaussian}.  

In the context of integrating Eqn.~\eqref{eq:diffequi} to obtain the chemical potential $\mu_{\alpha}(\tilde{\mathbf{x}})$, the gradient observations $\nabla \mathbf{Y}$ correspond to the rows $\mathbf{\Gamma}_{[\alpha,:]}$ evaluated at a set of independent variables $\tilde{\mathbf{x}} \in \mathbf{X}_g$. 
To fix the integration constant, at least one function value $Y_\mathrm{ref}$, at a reference composition $X_\mathrm{ref}$, is required. This reference can be chosen as a known or a defined baseline for $\mu_{\alpha}$, e.g., the chemical potential of the pure liquid of component $\alpha$ may be set to zero.
The GP then provides estimates of the function values $\mu_{\alpha}$ at new compositions $\tilde{\mathbf{x}} \in \mathbf{X}_*$ via Eqn.~\eqref{eq:GP}, along with corresponding uncertainty estimates from the diagonal elements of the covariance matrix in Eqn.~\eqref{eq:covarianceGP}.

\subsection{Active Learning}

Integrating over a multi-dimensional compositional space to accurately determine $\mu_\alpha(\tilde{\mathbf{x}})$ requires substantially more gradient data 
compared to the two-component case.
As such, careful selection of the composition points $\tilde{\mathbf{x}}$ to evaluate $\mathbf{\Gamma}$ becomes important.
Here we employ an active learning (AL) scheme to iteratively select additional compositions for performing more MD simulations.

During each AL round, to improve the accuracy of $\mu$ predicted from existing inputs $\mathbf{X}$, 
one selects $m$ additional compositions from a set of candidate points $\mathbf{X}_{g*}$, with $m$ being a user-defined number.
Intuitively, one wishes to sample points with large estimated gradient uncertainty, but not too close to each other.
Fig.~\ref{fig:GPandALschematic} illustrates such a scheme.

To this end, we first formulate a gradient covariance matrix $\mathbf{T}$ that combines the GP covariance for gradients $\mathrm{cov}(\nabla \tilde{\mathbf{Y}}_{*})$ in Eqn.~\eqref{eq:covarianceGP}.
The matrix element between a pair of compositions $ \tilde{\mathbf{x}}, \tilde{\mathbf{x}}'\in \mathbf{X}_{g*}$ is
\begin{multline}
      T (\tilde{\mathbf{x}}, \tilde{\mathbf{x}}') = 
      \sum_{\alpha=1}^C
      \mathrm{cov}\left( \nabla \mu_{\alpha}(\tilde{\mathbf{x}}),\nabla \mu_{\alpha}(\tilde{\mathbf{x}}') \right)\\
    =   \sum_{\alpha=1}^C
    \sum_{\beta=1}^f \sum_{\beta'=1}^f 
    \mathrm{cov}\left(
    \dfrac{\partial \mu_{\alpha} (\tilde{\mathbf{x}})}{\partial \tilde{x}_\beta},
    \dfrac{\partial \mu_{\alpha} (\tilde{\mathbf{x}}')}{\partial \tilde{x}'_{\beta'}}
     \right).
\end{multline} 

On the matrix $\mathbf{T}$, we then employ the CUR decomposition algorithm~\cite{michael2009CUR} to select $m$ columns as well as the same $m$ rows, to form matrices $\mathbf{C}$ and $\mathbf{R}$, respectively. 
These selected matrices form a low-rank approximation, $\mathbf{T} \approx \mathbf{C}\mathbf{U}\mathbf{R}$, where $\mathbf{U}$ is an $m\times m$ intersection matrix.
In other words, the CUR algorithm selects $m$ compositions whose correlations best represent the original covariance matrix.
In practice, the $m$ points are sequentially selected using a greedy deterministic algorithm, where the column with the highest statistical leverage score is picked. 
To avoid selecting similar points, the projection of $\mathbf{T}$ onto the selected column is removed before selecting the next one.

After that, additional MD simulations are performed for the $m$ selected compositions to collect the corresponding chemical potential derivatives.
The GP regression (Eqns.~\eqref{eq:GP} and~\eqref{eq:covarianceGP}) on the expanded data set then provides an updated estimate of the chemical potentials, chemical potential gradients, and their covariance matrices.

\section{Liquid Fe-Cu-Ni alloy mixing free energies}
\label{sec:FeCUNiLiAlloy}
\begin{figure*}
    \centering \includegraphics[width=0.8\linewidth]{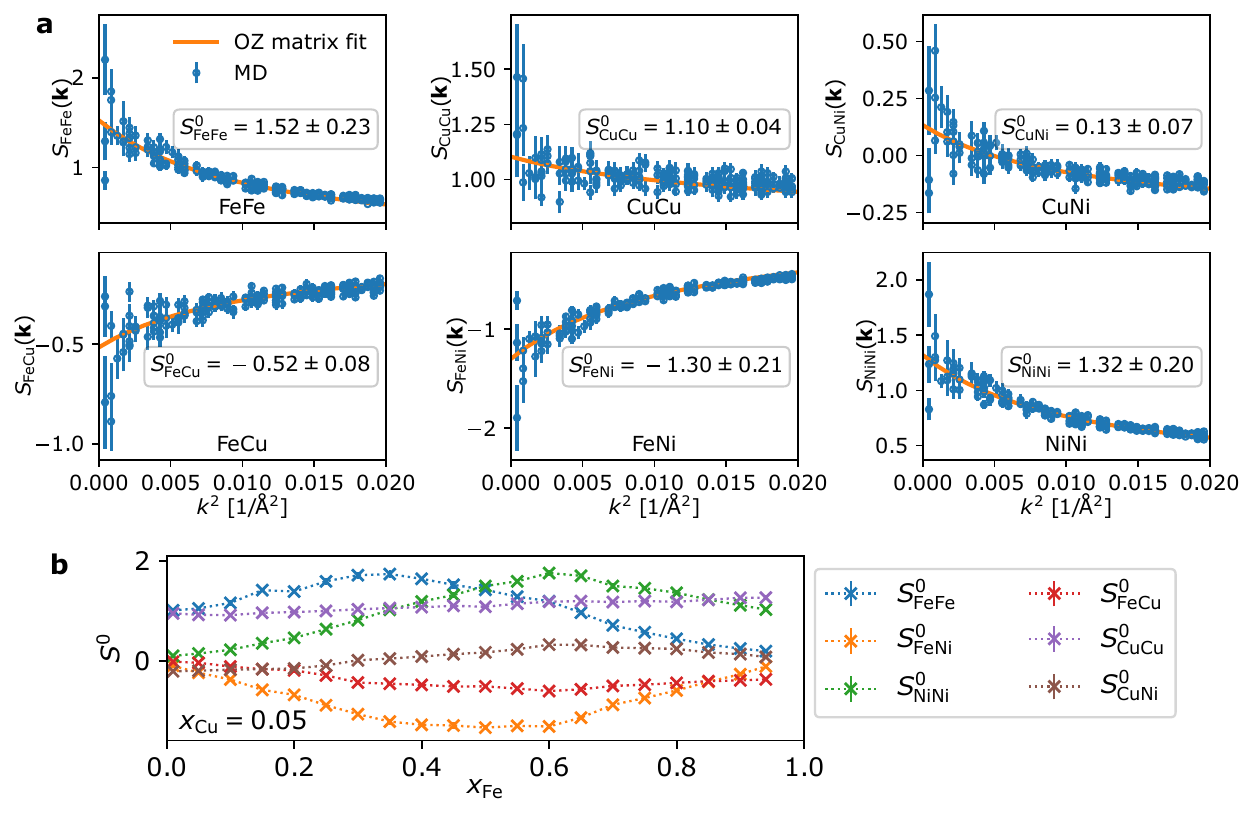} 
\caption{$S(\mathbf{k})$ and the extrapolated $S^0$ values for the Fe-Cu-Ni system.\\
\figLabelCapt{a} Computed $S(\mathbf{k})$ values for composition 
$(x_{\mathrm{Fe}},x_{\mathrm{Cu}},x_{\mathrm{Ni}})=(0.45,0.05,0.50)$ and the Ornstein-Zernike matrix fit.
\figLabelCapt{b} $S^0$ values for all component pairs as a function of
$x_{\mathrm{Fe}}$ at $x_{\mathrm{Cu}}=0.05$.
Error bars come from the statistical uncertainties in $S(\mathbf{k})$.}
\label{fig:Sk}
\end{figure*}

To benchmark the extended S0 method, we examine the liquid Fe-Cu-Ni alloy system, whose mixing Gibbs energies have been thoroughly studied in Ref.~\cite{computingTrinkle2025}, and have various implications for designing superalloys.

For systems with 8000 atoms and with different atomic fractions,
NPT simulations at 1~bar and 3000~K were performed in LAMMPS~\cite{LAMMPS}, using an embedded atom method (EAM) potential~\cite{ternaryBonny2009}.
The system was equilibrated for 0.5 ns followed by a 2 ns production run with a timestep of 1 fs, using an isotropic Nos\'{e}–Hoover barostat~\cite{shinoda2004rapid} and a stochastic velocity-rescaling thermostat~\cite{bussi2007Canonical}.

From each MD trajectory at a certain composition, we computed the structure factors between pairs of elements at different reciprocal-cell wavevectors $\mathbf{k}$ using Eqns.~\eqref{eq:sfsim} and~\eqref{eq:rhokt}. 
\figrefsub{fig:Sk}{a} shows an example of computed $S(\mathbf{k})$ values for composition 
$(x_{\mathrm{Fe}},x_{\mathrm{Cu}},x_{\mathrm{Ni}})=(0.45,0.05,0.5)$.
The orange curves are the matrix Ornstein-Zernike (OZ) fit (Eqn.~\eqref{eq:matrix_oz}) using a maximum cutoff of $k_{\mathrm{cut}}^2 = 0.005 \times 4\pi^2/\mathrm{\mathring{A}^2}$.
\figrefsub{fig:Sk}{b} shows the values $S^0$ for all component pairs as a function of $x_{\mathrm{Fe}}$ at the fixed $x_{\mathrm{Cu}}=0.05$.

 \begin{figure*}
    \centering \includegraphics[width=0.9\linewidth]{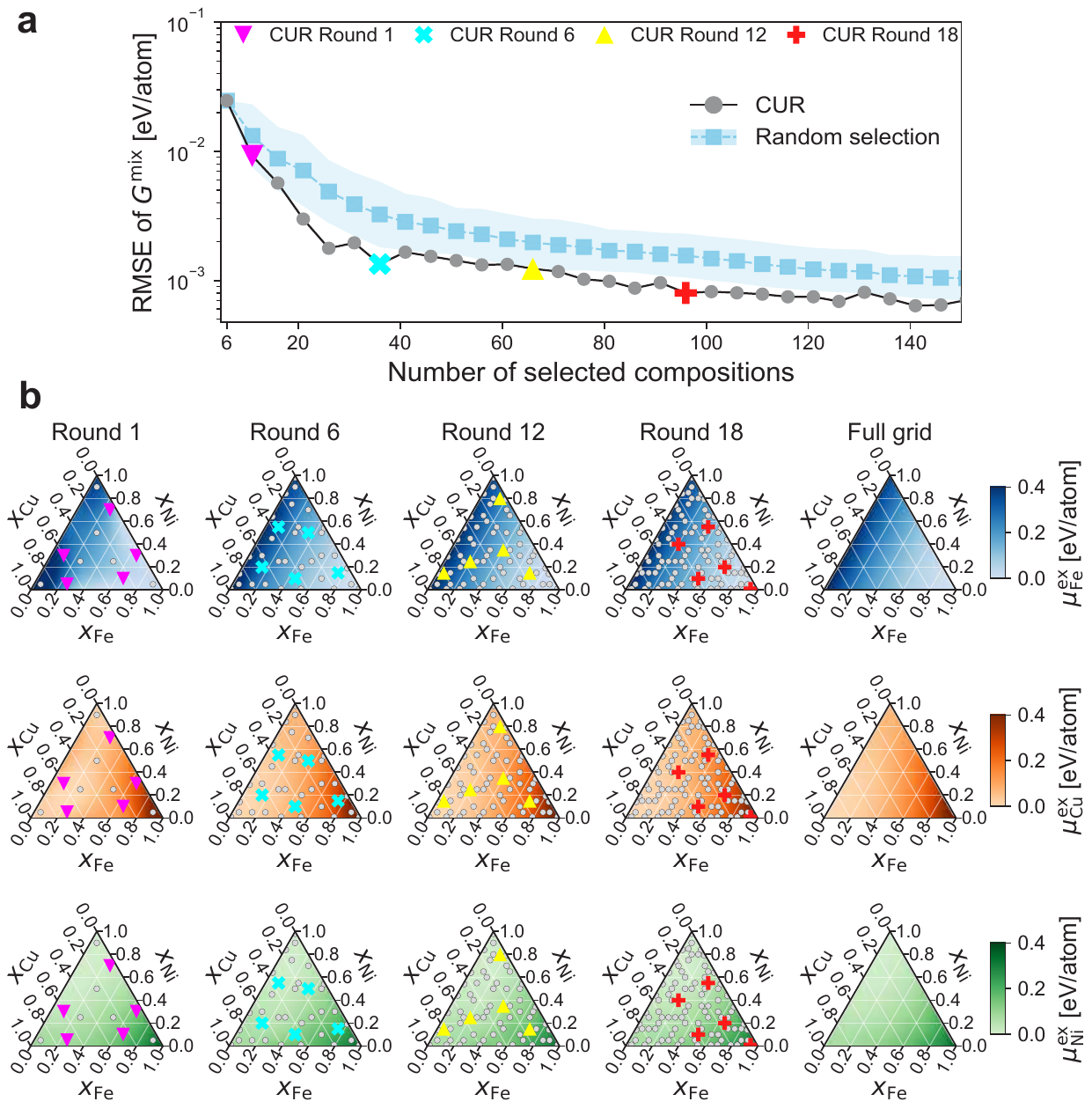} 
    \caption{A comparison of the chemical potential accuracy based on iteratively selecting new data points using CUR or random selection. \\
    \figLabelCapt{a} The comparison of the mixing free energy ($G^{\mathrm{mix}}$) root-mean-square errors (RMSEs) at each round, evaluated against the 231-point full grid reference. 
    \figLabelCapt{b} Evolutions of the excess chemical potential surfaces, $\mu^{\mathrm{ex}}_{i}$, of Fe (blue), Cu (orange), and Ni (green) as additional compositions are selected through successive CUR iterations. The reference chemical potential surfaces (rightmost column) were obtained from the full grid.}
    \label{fig:RandvsAL}
\end{figure*}

Using Eqns.~\eqref{eq:muder} and~\eqref{eq:k0lim}, the $S^0$ values can then be used to construct the matrix $\mathbf{U}$. 
To transform $\mathbf{U}$ into $\mathbf{\Gamma}$, Eqn.~\eqref{eq:gamma} was used with the independent atomic fractions set to be $\tilde{\mathbf{x}} = [ x_{\mathrm{Fe}},x_{\mathrm{Cu}}]^T$.
We then obtain $\mathbf{\Gamma}^{\mathrm{ex}}$ using Eqn.~\eqref{eq:gammaexeq},
and perform GP integration (Eqn.~\eqref{eq:GP}) to get the excess chemical potentials (Eqn.~\eqref{eq:mu_ex}).
To fix the integration constant, we set the excess chemical potential of the pure component to be zero, e.g., $\mu^{\mathrm{ex}}_{\mathrm{Fe}} (x_{\mathrm{Fe}}=1) = 0$~eV/atom, and include this value as a noiseless observation ($\sigma_f=0$~eV/atom) in the GP integration. 

We started the first round of active learning with the six initial compositions shown as gray circles in the first column of \figrefsub{fig:RandvsAL}{b}, and then iteratively expanded the data set using the CUR selection as illustrated in ~\figrefsub{fig:RandvsAL}{a}.
For the GP integration at each round, species-specific RBF length scales $\theta$ and gradient-noise level $\sigma_g$ were optimized by maximizing the log marginal likelihood. 

From the excess chemical potentials, the free energy of mixing, $G^{\mathrm{mix}}$, can be computed as
\begin{equation}
    G^{\mathrm{mix}} = \sum_{i}x_i \mu^{\mathrm{ex}}_i + k_{\rm B}T\sum_{i}x_{i}\ln{x_i}.
    \label{eq:mixfreeenergy}
\end{equation}

\figrefsub{fig:RandvsAL}{a} compares the root-mean-square errors (RMSEs) of $G^{\mathrm{mix}}$ from the CUR and random point selection strategies. 
The ground-truth $G^{\mathrm{mix}}$ as the reference here is calculated for the excess chemical potential computed with a dense grid with 231 points. 
The curve for random selection is aggregated from 100 independent runs, with the shaded band indicating the standard deviation.   
\figrefsub{fig:RandvsAL}{b} illustrates the evolution of the $\mu^{\mathrm{ex}}_{\mathrm{Fe}}$, $\mu^{\mathrm{ex}}_{\mathrm{Cu}}$ and $\mu^{\mathrm{ex}}_{\mathrm{Ni}}$ surfaces as CUR points are successively added. 
The CUR method typically exhibits a lower RMSE than the random point selection at a given sample size, suggesting its advantage in selecting informative data points in sampling the chemical potential gradients. 

\begin{figure}
    \centering \includegraphics[width=1.0\linewidth]{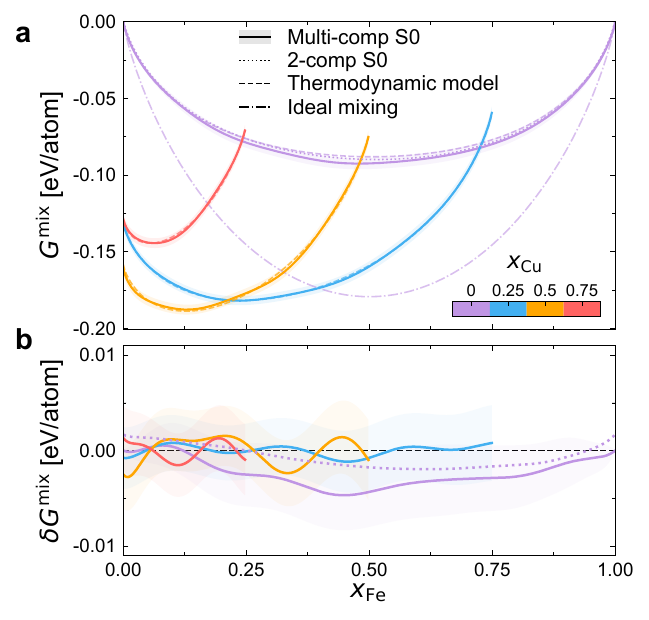} 
    \caption{\figLabelCapt{a} Free energy of mixing ($G^{\mathrm{mix}}$) for the Fe-Cu-Ni system at 3000 K as a function of $x_{\mathrm{Fe}}$ at fixed $x_{\mathrm{Cu}} = 0, 0.25, 0.5 ,0.75$, computed using the extended S0 method (solid line), the two-component S0 method (dotted line), and a previously reported thermodynamic model (dashed line)~\cite{computingTrinkle2025}, and compared to ideal $ G^{\mathrm{mix}}$ (dash-dotted line). The shaded bands denote the statistical uncertainties propagated from $S(\mathbf{k})$ values. \figLabelCapt{b} 
    Differences in the free energies of mixing relative to the thermodynamic model~\cite{computingTrinkle2025}: $\delta G^{\rm mix} = G^{\rm mix}$(Multi-comp S0) $- G^{\rm mix}$(ref) for the multi-component S0 method (solid lines) and $\delta G^{\rm mix} = G^{\rm mix}$(2-comp S0) $- G^{\rm mix}$(ref) for the two-component S0 method (dotted line).
}
    \label{fig:GmixMultiMethods}
\end{figure}

To validate the computed chemical potentials, \figrefsub{fig:GmixMultiMethods}{a} compares the computed $G^{\mathrm{mix}}$ curves at fixed $x_{\mathrm{Cu}}= 0, 0.25, 0.5, 0.75$ with a previous CALPHAD-style thermodynamic model parameterized by chemical potential differences between species that were computed using the virtual semigrand canonical Widom approach~\cite{computingTrinkle2025}. 
Moreover, \figrefsub{fig:GmixMultiMethods}{b} shows that the absolute
difference in $G^{\mathrm{mix}}$ between the multi-component S0
method and the thermodynamic model remains below 0.005~eV/atom across all compositions.
For $x_{\mathrm{Cu}}=0$, we also show $G^{\mathrm{mix}}$ computed with the two-component S0 method based on a dense $x_{\mathrm{Fe}}$ grid with a spacing smaller than 0.03.
Overall, all three methods have excellent agreement for this fully miscible and non-ideal mixture.

\section{Paracetamol Solubility in Water-Ethanol Mixtures}
Accurately predicting drug solubility in mixed solvents is essential for pharmaceutical development, particularly for the design of crystallization-based manufacturing and purification processes~\cite{variankaval2008form,ruether2009modeling,tung2023crystallization}. 
Here we apply the extended S0 framework to the paracetamol–water–ethanol ternary system, which combines a prototypical active pharmaceutical ingredient with a widely used pharmaceutical co-solvent. 

We performed MD simulations at 303.15~K and 1~atm using LAMMPS~\cite{LAMMPS} for 262 solution compositions, spanning mole fractions of paracetamol ($x_{\rm para}$) ranging from $5\times10^{-4}$ to 0.2. 
Each simulation cell contained about 10,000 solvent molecules with the solute-free ethanol mole fraction ranging from 0 to 1. 
The system was described using the CHARMM36~\cite{CHARMM36} parameterization as specified in Ref.~\cite{reinhardt2023streamlined}.
Each system was equilibrated for 3~ns, followed by a 3~ns production run with a 2~fs timestep, and snapshots were saved every 2~ps.
The system temperature and pressure were controlled using a Nos\'{e}–Hoover thermostat and barostat~\cite{shinoda2004rapid}, respectively.

The partial structure-factor matrix, $\mathbf{S}(\mathbf{k})$, was calculated using Eqns.~\eqref{eq:sfsim} and~\eqref{eq:rhokt} from sampled NPT trajectories, where water, paracetamol, and ethanol were represented by the oxygen, nitrogen, and hydroxyl-adjacent carbon atoms, respectively. 
The $S^0$ values were fitted using Eqn.~\eqref{eq:matrix_oz} with $k^2_{\rm cut}=0.005 \times 4\pi^2/\AA^2$. The resulting $S^0$ matrix was then used to compute the excess chemical potential derivative matrix, $\mathbf{\Gamma}^{\rm ex}$, using Eqns.~\eqref{eq:k0lim}, \eqref{eq:muder}, \eqref{eq:gamma}, and \eqref{eq:gammaexeq}, selecting the independent mole fractions $\tilde{\mathbf{x}}=[x_{\rm para},x_{\rm eth}]^T$.

$\mathbf{\Gamma}^{\rm ex}$ was then GP-integrated to obtain the excess chemical potential of paracetamol, $\mu^{\rm ex}_{\rm para}$, 
which is related to the absolute chemical potential by
\begin{equation}
\mu_{\rm para}=k_{\rm B}T\ln(x_{\rm para})+\mu^{\rm ex}_{\rm para}(x_{\rm para},x_{\rm eth}),
\label{eq:totalchempotpara}
\end{equation}
with the ideal contribution depending only on $x_{\rm para}$.
To fix the integration constant of the GP, we used two absolute chemical potentials of paracetamol at 303.15~K computed using a combination of the free-energy perturbation (FEP) and thermodynamic integration (TI) methods from Ref.~\cite{reinhardt2023streamlined}: 
$\mu^{\rm ex}_{\rm para}(x_{\rm para}=0.0084)=-22.99\pm0.38$~kJ/mol in pure water, and
$\mu^{\rm ex}_{\rm para}(x_{\rm para}=0.0135)=-66.36\pm0.14$~kJ/mol in pure ethanol.
We used the input-warped RBF GP model defined in Eqn.~\eqref{eq:input_warp}, which makes the effective length scale composition-dependent at the cost of one additional parameter per warped dimension.
In particular, we applied $\boldsymbol{\alpha}=(0,0.125)$ and $\boldsymbol{\theta}=(0.12, 0.13)$ along
$(x_{\rm para},x_{\rm eth})$, with uniform gradient noise $\sigma_g=0.10$~kJ/mol.

\begin{figure}[htbp!]
    \centering \includegraphics[width=0.95\linewidth]{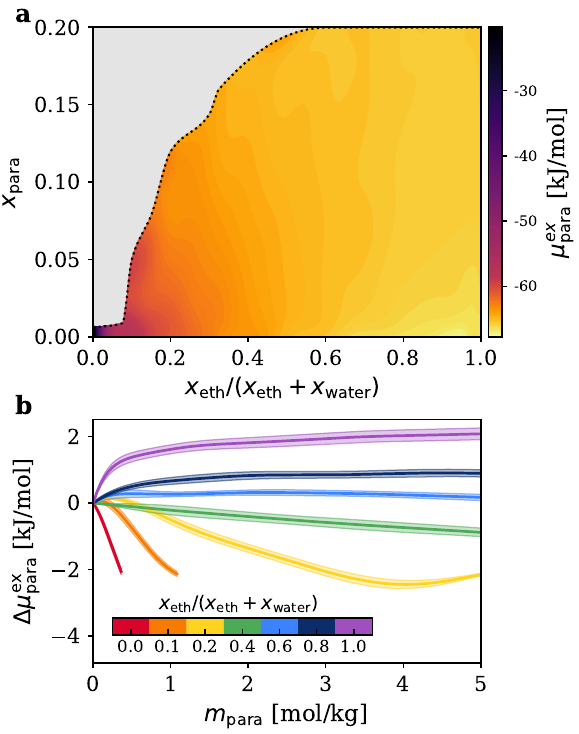} 
    \caption{Excess chemical potential of paracetamol, $\mu^{\rm ex}_{\rm para}$, in water--ethanol mixtures at 303.15~K and 1~atm.
    \figLabelCapt{a} $\mu^{\rm ex}_{\rm para}$ as a function of the paracetamol mole fraction, $x_{\rm para}$, and the solute-free ethanol mole fraction, $x_{\rm eth}/(x_{\rm eth}+x_{\rm water})$, computed from the GP integration of $\mathbf{\Gamma}^{\rm ex}$.
    \figLabelCapt{b} Change in the excess chemical potential relative to the infinite-dilution limit,
    $\Delta\mu^{\rm ex}_{\rm para}
    =\mu^{\rm ex}_{\rm para}(x_{\rm para},x_{\rm eth})
    -\mu^{\rm ex}_{\rm para}(0,x_{\rm eth})$,
    as a function of paracetamol molality, $m_{\rm para}$, at fixed solute-free ethanol mole fractions $x_{\rm eth}/(x_{\rm eth}+x_{\rm water})$ =0, 0.1, 0.2, 0.4, 0.6, 0.8, and 1. The shaded regions are the statistical uncertainties propagated from $S(\mathbf{k})$ values.}
    \label{fig:muex_para}
\end{figure}

\figrefsub{fig:muex_para}{a} shows the $\mu^{\rm ex}_{\rm para}$ determined by the GP. The dotted line depicts the domain of compositions in the chemical potential integration, which exceeds the solubility limit for paracetamol for all ethanol fractions. \figrefsub{fig:muex_para}{a} shows that $\mu^{\rm ex}_{\rm para}$ decreases rapidly with increasing ethanol fraction, while showing less sensitivity to $x_{\rm para}$.
\figrefsub{fig:muex_para}{b} further shows quantitatively the concentration-induced change in the excess chemical potential, $\Delta\mu^{\rm ex}_{\rm para} = \mu^{\rm ex}_{\rm para}(x_{\rm para},x_{\rm eth}) - \mu^{\rm ex}_{\rm para}(0,x_{\rm eth})$, at different solvent compositions.
In water-rich solvents, $\Delta\mu^{\rm ex}_{\rm para}$ decreases steeply with paracetamol molality $m_{\rm para}$.
As the ethanol content increases, this concentration dependence progressively weakens and eventually reverses sign, with $\Delta\mu^{\rm ex}_{\rm para}$ becoming positive.

\begin{figure}[htbp!]
    \centering \includegraphics[width=0.95\linewidth]{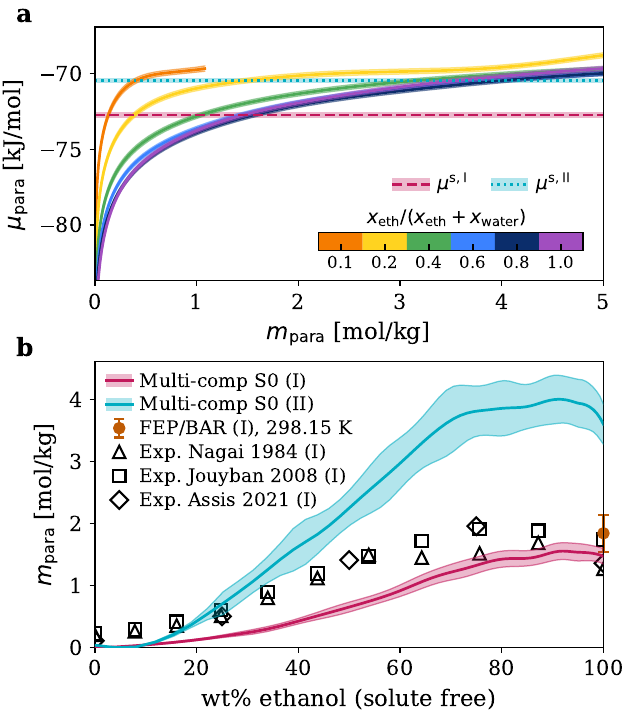}
    \caption{Solubility of paracetamol at 303.15~K and 1~atm.
    \figLabelCapt{a} Chemical potential of solvated paracetamol at fixed solute-free ethanol mole fraction $x_{\rm eth}/(x_{\rm eth}+x_{\rm water})$, with curve widths indicating the statistical uncertainties.
    The horizontal dashed red and dotted blue lines denote the chemical potentials of crystalline polymorph Forms~I and II reported in Ref.~\cite{reinhardt2023streamlined}, respectively, with shaded bands denoting their uncertainties. 
    \figLabelCapt{b} Solubilities of paracetamol Forms~I and II predicted using the multi-component S0 method (solid curves) as functions of the solute-free ethanol weight percentage (wt\%).
    Experimental Form~I data at 303.15~K and 1~atm are shown as black triangles~\cite{NAGAISolubility1984}, squares~\cite{JouybanSolubility2008}, and diamonds~\cite{Gabriel2021Solid}. 
    The orange circle shows the previous FEP/BAR estimate for Form~I in pure ethanol at 298.15~K and 1~bar~\cite{Bellucci2019Solubulity}.}
\label{fig:sol_para}
\end{figure}

Next, we use the chemical potential of solvated paracetamol $\mu_{\rm para}$ (~\figrefsub{fig:sol_para}{a}) to determine the solubilities for
two crystalline polymorphs: Form~I~\cite{ou2020general} and Form~II~\cite{thomas2011paracetamol}. 
As indicated by the horizontal lines, their solid-state chemical potentials at 303.15~K and 1~atm ($\mu^{\rm s, I}=-72.74\pm0.17$~kJ/mol for form I and $\mu^{\rm s, II}=-70.46\pm0.17$~kJ/mol for form II) were taken from previous 
thermodynamic integration calculations~\cite{reinhardt2023streamlined}.  
In~\figrefsub{fig:sol_para}{a}, the intersections of $\mu_{\rm para}(m_{\rm para})$ at different solvent compositions with $\mu^{\rm s, I}$ and $\mu^{\rm s, II}$ determine their corresponding solubilities.

\figrefsub{fig:sol_para}{b} shows the solubilities of form I and form II in solvents with different ethanol weight percentages (wt\%) on a solute-free basis.
In particular, form I solubility in pure ethanol is $1.46\pm0.13$~mol/kg,
which falls within the spread of the experimental values~\cite{JouybanSolubility2008,NAGAISolubility1984,Gabriel2021Solid},
as well as previous FEP with Bennett acceptance ratio (BAR) calculations~\cite{Bellucci2019Solubulity} based on a similar CHARMM-family force-field. 
In contrast, the predicted solubilities of Forms~I and II in pure water are well below the experimental range~\cite{JouybanSolubility2008,NAGAISolubility1984,Gabriel2021Solid,nishigaki2021growth}.
A comparable discrepancy was reported previously~\cite{reinhardt2023streamlined}, and was attributed to the limitation of the empirical force field employed.

The S0-method derived solubilities 
of both polymorphs increase substantially with ethanol content and approach a plateau in the ethanol-rich end. 
This trend is qualitatively consistent with the three experimental measurements of Form~I~\cite{JouybanSolubility2008,NAGAISolubility1984,Gabriel2021Solid}, which place their sampled maxima between 75 and 90~wt\% ethanol (solute-free basis), as shown in \figrefsub{fig:sol_para}{b}.
This suggests that the S0 method is able to capture the subtle chemical potential dependence on solute and mixed solvent interactions.

\section{Conclusions}

We present the extended S0 method which computes the chemical potentials of multi-component mixtures from equilibrium NPT MD simulations. 
Using structure factors computed from simulation, we can extract the derivatives of chemical potentials with respect to particle number fractions.
This is thermodynamically equivalent to the integration along molar concentrations in the original S0 method~\cite{Cheng2022Computing},
and perhaps a bit more straightforward in implementation.

Gaussian process regression is then utilized to integrate these multi-dimensional derivatives of chemical potentials with respect to particle number fractions.
This again differs from the numerical integration scheme in the original paper~\cite{Cheng2022Computing}.
Being more data efficient and robust, GP becomes advantageous for the integration in the multi-dimensional space.
It is straightforward when more than one absolute chemical potential reference point is included in the integration.
Another advantage of the GP is that,
from the predicted covariance, one can use active learning to select the most informative set of compositions to simulate and iteratively improve the chemical potential estimations. 
The GP integration and active learning loops can also be transferable to approaches based on Kirkwood–Buff integrals~\cite{Kang2008Kirkwood, Simon2022Kirkwood, robustBusselez2025}.

In this paper, we demonstrate the S0 method on two three-component systems, but the method can also be applied to more complex systems with even more components,
although in such cases more simulations at different compositions may be needed for integrating the chemical potentials.
The Gaussian process regression and active learning combination can help mitigate such computational expense and make integration more numerically tractable.
The resulting composition-dependent chemical potentials could also
support future studies of nonadditive effects in mixed-electrolyte solutions~\cite{Moghaddam2015Hofmeister,Bruce2018Nonadditive,MandalaparthyInfluence2022,Mandalaparthy2026Beyond} and atmospheric aerosols~\cite{Dutcher2012Statistical,Dutcher2013Thermodynamic}.

Both applications considered here only involve neutral components.
Treating explicitly charged particles requires an extension of the present formulation, 
which we will explore in a future manuscript, Part II: Charged multicomponent mixtures. 
For neutral multicomponent systems, the only intrinsic constraint on composition is normalization, whereas charged systems must additionally satisfy charge neutrality. 
Moreover, the long-range Coulomb interaction changes the small-wavenumber behavior of the partial structure factors: 
charge-density fluctuations are suppressed in the $k\rightarrow 0$ limit, so the Ornstein–Zernike extrapolation (Eqn.~\eqref{eq:oz}) used here must be replaced by a charge-aware treatment. 

To conclude, we generalize the S0 method to 
neutral multi-component systems.
This extends the usage of the method as a generally applicable and easy-to-use framework for computing chemical potentials in complex mixtures.

\section{Appendix}

\subsection{Ensemble switching}
\setcounter{equation}{0}
\renewcommand{\theequation}{A\arabic{equation}}

\label{sec:ensembleSwitching}
The matrix $\mathbf{A}$ with entries
\begin{equation}
        A_{\alpha \beta} = \dfrac{N}{k_{\rm B}T}\left(\dfrac{\partial\mu_{\alpha}}{\partial  n_{\beta} }\right)_{T, V} 
        \label{eq:A}
\end{equation}
is defined in the canonical ensemble.
One can show that $\mathbf{A}=\mathbf{B}^{-1}$,
as at the thermodynamic limit $n_i = \langle n_{i} \rangle_{\{\mu\},V,T}$ and
\begin{equation}
\delta_{\alpha \beta} = 
\dfrac{\partial \mu_{\alpha}}{\partial \mu_{\beta}} =\sum_{i}\dfrac{\partial \mu_{\alpha}}{\partial n_{i}}\dfrac{\partial n_i}{\partial \mu_{\beta}}  =\sum_{i} A_{\alpha i} B_{i \beta}. 
\label{eq:ident}
\end{equation}

One can write $\mathbf{U}$ in terms of $\mathbf{A}$ as 
\begin{equation}
\mathbf{U}= k_{\rm B}T\mathbf{A} - \dfrac{N}{V\kappa_{T}}  \mathbf{v}\mathbf{v}^{T},
\label{eq:Thermidsimp}
\end{equation}
where $\kappa_{T} = -\dfrac{1}{V}\left(\partial V/\partial P\right)_{T,\{n\}}$, and $\textbf{v}$ is a column vector containing the partial molar volume of each component, $v_i = \left(\partial V /\partial n_{i}\right)_{T, P}$. 
The derivation of Eqn.~\eqref{eq:Thermidsimp} starts with the thermodynamic identity
\begin{multline}
 \left(\dfrac{\partial \mu_{\alpha}}{\partial n_{\beta}}\right)_{T, V, n \neq n_\beta }  = \left(\dfrac{\partial \mu_{\alpha}}{\partial n_{\beta}}\right)_{T, P, n \neq n_\beta } \\  +\left(\dfrac{\partial \mu_{\alpha}}{\partial P}\right)_{T, \{n\}} 
\left(\dfrac{\partial P}{\partial n_{\beta}}\right)_{T, V, n \neq n_\beta } ,
\label{eq:thermoiden}
\end{multline}
and then we use a Maxwell relation to replace
\begin{equation}
 \left(\dfrac{\partial \mu_{\alpha}}{\partial P}\right)_{T, \{n\}}  = 
\left(\dfrac{\partial V}{\partial n_{\alpha}}\right)_{T, P} = v_\alpha,
\end{equation}
and substitute $(\partial P /\partial n_{\beta})_{T, V}$ using the cyclic chain rule
\begin{equation}
\left(\dfrac{\partial P}{\partial n_{\beta}}\right)_{T, V}  \left(\dfrac{\partial n_{\beta}}{\partial V}\right)_{T, P} \left(\dfrac{\partial V}{\partial P}\right)_{T, \{n\}}  = -1. 
\end{equation}

To further eliminate $\mathbf{v}$ and $\kappa_{T}$ from Eqn.~\eqref{eq:Thermidsimp}, 
we left multiply it by $\mathbf{x}^{T}$, 
\begin{equation}
   \mathbf{x}^{T}\mathbf{U}= k_{\rm B}T\mathbf{x}^{T}\mathbf{A} - \dfrac{N}{V\kappa_{T}}(\mathbf{x}^{T}\mathbf{v})\mathbf{v}^{T}.
   \label{eqn:xU}
\end{equation}
We use the Gibbs-Duhem equation
\begin{equation}
\mathbf{x}^T\mathbf{U} = \mathbf{0}^T, 
\label{eq:gibbsDuhem}
\end{equation}
and the volume relationship $\mathbf{x}^T \mathbf{v}=\dfrac{V}{N}$,
to simplify Eqn.~\eqref{eqn:xU} into
\begin{equation}
    \dfrac{\mathbf{v}^{T}}{\kappa_{T}} = k_{\rm B} T\mathbf{x}^T\mathbf{A}.
    \label{eq:Vtran}
\end{equation}
Then we right multiply Eqn.~\eqref{eq:Vtran} by $\mathbf{x}$ yielding the expression for $\kappa_T$,
\begin{equation}
    \kappa_{T} = \dfrac{V}{N k_{\rm B}T} \dfrac{1}{\mathbf{x}^T\mathbf{A}\mathbf{x}}.
    \label{eqn:kappaT}
\end{equation}
Substituting the term $\mathbf{v}\mathbf{v}^T/\kappa_{T}$ from Eqn.~\eqref{eq:Thermidsimp} by combining Eqn.~\eqref{eq:Vtran} and Eqn.~\eqref{eqn:kappaT},
and replacing $\mathbf{A}$ with $\mathbf{B}^{-1}$ finally leads to Eqn.~\eqref{eq:muder}.

\textbf{Acknowledgments}
Funding acknowledgement: Research reported in this publication was supported by the National Institute of General Medical Sciences of the National Institutes of Health under Award Number R35GM159986.
R.S., M.A. and B.C. were supported by the Laboratory Directed Research and Development (LDRD) Program at Lawrence Berkeley National Laboratory (LBNL) under project ID 110990-001.
The authors acknowledge the research computing facilities provided by BRC UCB and LRC Lawrence Berkeley National Laboratory. 

\textbf{Data availability statement}
Empirical force-field parameters, MD input scripts, MD results, and data-analysis scripts generated for this study are available in the public repository: \url{https://github.com/ChengUCB/extended_S0}.

\textbf{Code availability} 
The Gaussian process regression code is publicly available at \url{https://github.com/ChengUCB/GPR_grad}. The multi-component S0 code is publicly available at \url{https://github.com/ChengUCB/S0_multi/}.

\end{document}